# Role of water in the enzymatic catalysis: study of ATP + AMP → 2ADP conversion by adenylate kinase

Bharat V. Adkar, Biman Jana and Biman Bagchi*

Solid State and Structural Chemistry Unit

Indian Institute of Science, Bangalore 560012, India

## Abstract

The catalytic conversion ATP + AMP → 2ADP by the enzyme adenylate kinase (ADK) involves the binding of one ATP molecule to the LID domain and one AMP molecule to the NMP domain. The latter is followed by a phosphate transfer, and then the release of two ADP molecules. We have computed a novel two dimensional configurational free energy surface (2DCFES), with one reaction coordinate each for the LID and the NMP domain motions, with explicit interactions with water. Our computed 2DCFES clearly reveals the existence of a stable *half-open-half-closed (HOHC) intermediate state* of the enzyme. Cycling of the enzyme through the HOHC state reduces the conformational free energy barrier for the reaction by about 20 kJ/mol. We find that the stability of the *half-open-half-closed* state (missed in all earlier studies with implicit solvent model) is largely because of the increase of specific interactions of the polar amino acid side chains with water, particularly with the arginine and the histidine residues. F*ree energy surface of the LID domain is rather rugged,* which can conveniently slow down LID's conformational motion, thus facilitating a new substrate capture after the product release in the catalytic cycle.

# I. Introduction

Understanding of the key factors that control the kinetics of enzyme catalysis has been a long standing goal of chemists and biochemists [1-3]. The explicit role of water in controlling the enzyme kinetics and the pathway has often been discussed in the enzyme literature [4-8]. It is of course well-known that water plays a crucial role in determining the structure and function of biomolecules. In a series of landmark papers, Fleming and co-workers have used three-pulse photon echo peak shift (3PEPS) and time-resolved fluorescence depolarization techniques to measure the magnitudes and time scales of energy fluctuations of the chromophore eosin bound to lysozyme in aqueous solution at physiological temperature [9-11]. They have examined the influence of inhibitor binding on the internal motion of lysozyme [9-11]. In a separate study, Fleming and co-workers examined the role of water on the residue motion of a protein using molecular dynamics simulation [12]. The important finding of this study is that in the presence of water the dynamics becomes slower [12]. Theese studies partly motivated us to study the role of water in enzyme catalysis.

While the specific role of water in the chemical step of enzymatic reactions has been explored in great details by elegant papers of Warshel and co-workers [13-14], the explicit role of water in the conformational cycling step of enzyme catalysis has hardly been explored. Many of the earlier studies were carried out with an implicit solvent model (where water is replaced by a dielectric continuum) in this context. Such studies have only a limited scope to understand the role of water on the conformational fluctuation because they ignore much of specific interactions (such as hydrogen bonding) that a water molecule is capable of. The molecular dynamics study of ligand-induced flap closing in HIV-1 protease demonstrates that the structural water plays a critical role in flap closing dynamics by destabilizing the hydrophobic clusters and subsequently by mediating the flap−ligand interactions [15].

Catalysis of the reaction, ATP + AMP → 2ADP, by the enzyme adenylate kinase (ADK) is a subject of tremendous current interest [16-23]. From the experimental kinetic data, the estimated rate of the catalysis for this reaction is 263 s$^{-1}$ which corresponds to a barrier height of 55 – 60 kJ/mol. While may of the studies have concluded that the barrier is associated with conformational fluctuation of the ADK [16-22], recent study by Warshel and co-workers finds that the activation barrier of ADK catalysis is primarily intrinsic due to the chemical step based on their full EVB model [23]. Thus, whether this barrier is due to the conformational fluctuation or due to the intrinsic phosphate transfer step is not yet clear. Recent experimental results by Kern and coworkers have *implied* the existence of an intermediate, half-open-half-closed (HOHC) state of ADK. However, the important issue of the mode of involvement of this intermediate state in determining the activation barrier and also in the enzyme catalytic cycle was left unquestioned and unanswered [16-18]. Specific characterization and explicit demonstration of the role of this intermediate state were also not carried out. Pulse-EPR spectroscopy study of the unliganded HIV-1 protease enzyme implies the existence of an intermediate "semiopen/semiopen" conformer which makes the understanding of the catalytic cycle significantly better [24].

There are several interesting issues about the coupling between the enzymatic fluctuation ({Q} coordinates) and the intrinsic reaction (X coordinate) of the chemical reaction. Warshel and coworkers [23] have examined whether there is a flow of energy from the {Q} coordinates to the X coordinate (and vice versa) during the cycle. They argued that enzymatic fluctuation does not catalyze the chemical step [23]. *However, in addition to the coupling of enzyme fluctuations to*

*the chemical step, there remains the important issue to the role of enzyme fluctuations leading to the reaction geometry.*

In a series of recent papers, Min, Xie and Bagchi discussed how such apparently non-reactive enzymatic fluctuation can help the intrinsic chemical step ***without any direct flow of energy*** [25-26]. The enzymatic fluctuations may be required to bring the reactant to a configuration where the chemical step can occur. If that particular reaction configuration is unique, then the rate of the catalysis (k) can be written as $\frac{1}{k} = \frac{1}{k_{Conf}} + \frac{1}{k_{Chem}} + \frac{1}{k_{S-bind}}$, where $k_{Conf}$ is the rate of enzyme fluctuation to the unique geometry, $k_{Chem}$ is the contribution from the intrinsic chemical step and $k_{S-bind}$ is the rate of the substrate binding to free enzyme. Min *et al.* showed how the rate constants ($k_{cat}$ and $K_M$) in Michaelis–Menten (MM) (in original MM equation, $k_{cat}$ is only a function of $k_{Chem}$ and $K_M$ is a function of $k_{Chem}$ and substrate concentration) equation get modified by fast or slow rate of enzymatic conformational fluctuation [26]. Importantly, they also discussed how an enzyme can operate in a non-equilibrium steady state to enhance the rate of catalysis [25-26]. These authors further pointed out that by staying in a non-equilibrium steady state, the enzyme can minimize the free energy barrier for conformational displacement [25-26]. Paradoxically, operating from this non-equilibrium steady state may actually lead to a *reduction* of the role of conformational fluctuations in catalysis. However, this Min-Xie-Bagchi theory assumes the existence of an intermediate state in the relaxation of the enzyme conformation *after the product release and prior to another substrate capture*. For this scenario to be applicable, this intermediate state needs to have relatively long lifetime and certain degree of stability so that a new substrate capture can occur after product release. We show below that the catalysis of ADK nicely fits into this picture.

We use very long computer simulations (with explicit water) to obtain an appropriate two dimensional free energy surface (2DCFES), described below, and carry out theoretical analysis to explore the existence of the intermediate state. We observe an intermediate HOHC state of ADK during the catalytic cycle and explicitly characterize the same. Importantly, cycling of the enzyme through the HOHC state is found to reduce the conformational free energy barrier by about 20 kJ/mol. We find that the specific interactions of water with the polar amino acid residues stabilize the HOHC intermediate state of the enzyme --- earlier simulation studies without explicit water failed to observe this state. We also discuss about the involvement of this intermediate enzymatic state in the catalytic cycle. In our present study, the calculated 2DCFES describe the conformational fluctuation of the enzyme in ligand free condition and we suppose that the nature of the fluctuation does not depend on the presence of ligand. We also like to point out that the present work does not concentrate on the barrier of intrinsic chemical step and it has been calculated explicitly elsewhere [23].

## II. Configurational Reaction Coordinates (CRC)

ADK undergoes large amplitude motions during catalysis. Both LID and the NMP domains open by a considerable amount to allow the ATP and AMP to get in, then close for the reaction and then again open to let the product go out. A key step in the catalysis (in addition, of course, to the chemical reaction step of phosphate transfer) can be another substrate capture after the product release as the continuation of the cycle is important. We have examined the conformational fluctuation of ADK by following the evolution of root mean square deviation (RMSD). We have also calculated the center of mass distances between LID/NMP and CORE domains and the inter-domain hinge angles. The result for the variation of RMSD, inter-domain distance and hinge angle involving the LID domain is shown in **Figure 1.** The existence of the intermediate state of the ADK (HOHC) is evident in all the quantities plotted in figure 1 (the elliptic region in the middle). The centre of mass distance between the LID and CORE domains in the HOHC state is found to be about 26 Å. The analysis also suggests that the conformational fluctuation in ADK is mainly due to the inter-domain displacement and not due to the change in structure of corresponding domains as the time evolution of these two quantities is very similar.

**Figure 1**

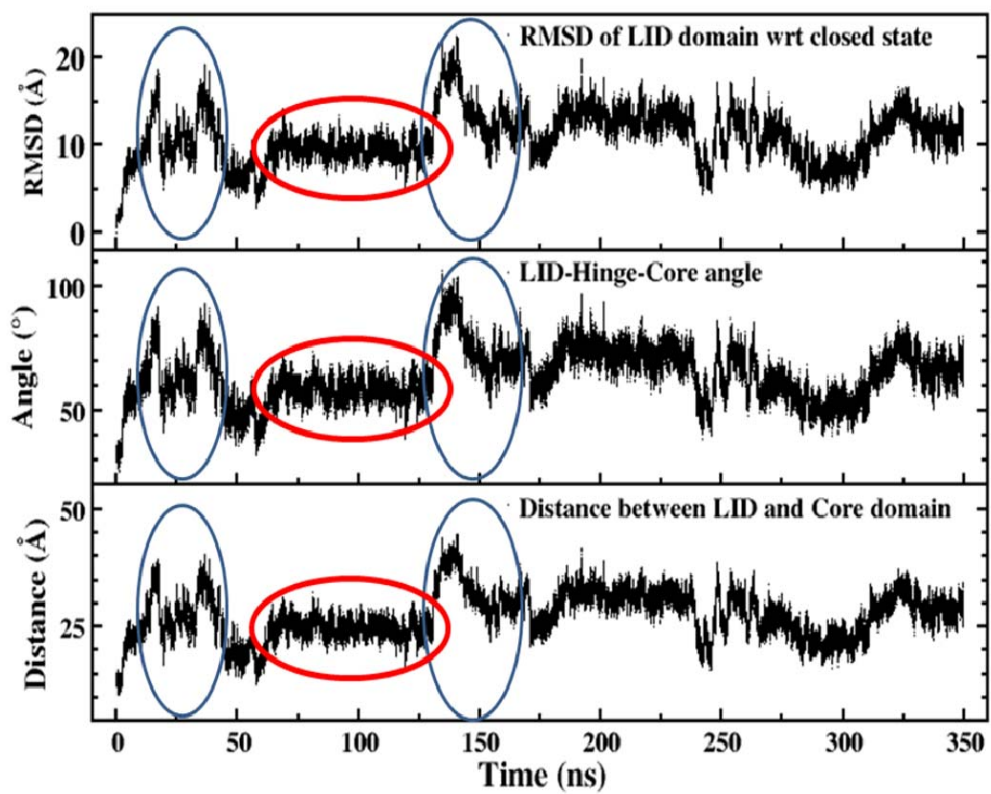

Therefore, we choose the centre of mass distances between LID and CORE ($R^{CM}_{LID-CORE}$) and NMP and CORE ($R^{CM}_{NMP-CORE}$) as the two order parameters for the free energy surface calculation. These serve as our configurational reaction coordinates.

# III. The two-dimensional configurational free energy surface (2DCFES) and conformational activation barriers

The free energy surface of the corresponding domain motions of the ADK is calculated by using umbrella sampling (values of the parameter used are provided in the **Method** section. The simulations involved more than 80,000 water molecules for the free energy calculations and span more than 20 μs). The free energy surface has been obtained for the LID domain motion at *13 distances* of NMP-CORE and for the NMP domain motion at *27 distances* of LID-CORE.

The configurational free energy surface, 2DCFES, of the free enzyme, as a function of CRC, is presented in **Figure 2a**. The conformation of ADK at $R^{CM}_{LID-CORE} \approx 26$ Å and $R^{CM}_{NMP-CORE} \approx 19$ Å forms a stable minimum for the free enzyme. This conformation belongs neither to the fully open state nor to the fully closed state ensemble. We have termed this ensemble of stable intermediate conformation as *half-open-half-closed (HOHC)* state hereafter. The presence of such a stable intermediate state is further confirmed by the trajectory analysis (see **Figure 1**).

The computed 2DCFES of the free enzyme reveals that the HOHC state is indeed a stable conformation of the ADK in solution. This state is separated by a barrier of about 60 kJ/mol from the fully open state ($R^{CM}_{LID-CORE} \approx 29$ Å and $R^{CM}_{NMP-CORE} \approx 21$ Å) and this corresponds to a timescale for conformational fluctuation between the open and the HOHC states of the order of milliseconds. This is consistent with the timescale ($\sim 286$ s$^{-1}$) observed in experiment [16-18]. However, in the presence of the ligand, the closed state ($R^{CM}_{LID-CORE} \approx 20.5$ Å and $R^{CM}_{NMP-CORE} \approx 18.3$ Å) is the stable state in the free energy surface. Thus, in a continuous catalytic cycle, ADK may shuttle between the HOHC and the closed states, rather than between the fully open and the closed state.

If we assume that the barrier of conformational fluctuation, away from the closed state, does not depend significantly on ligand binding, the conformational barrier ADK has to surmount from the HOHC state to reach near the closed state is about 40 kJ/mol. This barrier height is much lower than that between fully open and fully closed states and timescale associated with this conformational cycling is also much smaller. In such a situation, the catalytic cycle of the ADK gets modified (will be discussed later). Thus, the presence of stable intermediate state of ADK (HOHC) reduces the conformational fluctuation barrier related to catalytic cycle by about 20 kJ/mol. An additional outcome of our analysis is that the phosphate transfer might control the rate of the reaction.

**Figure 2(a)**

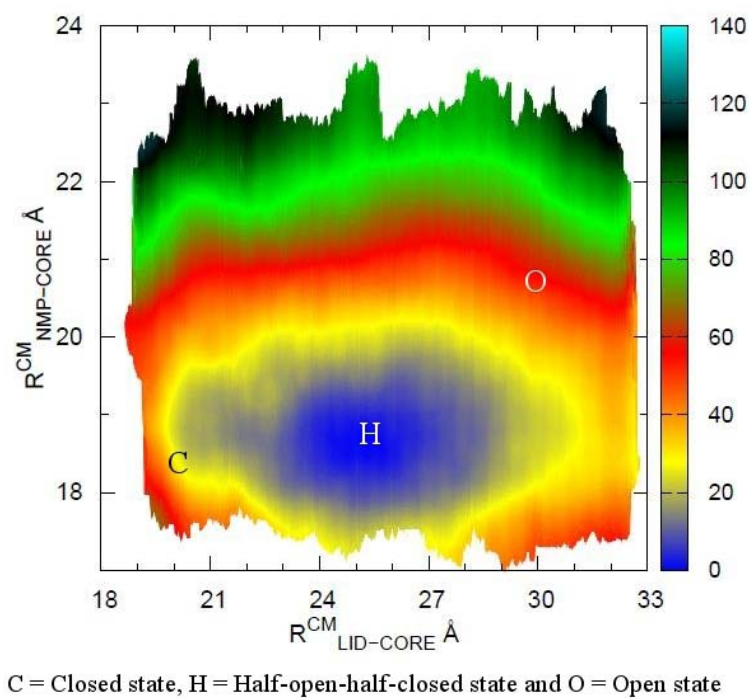

C = Closed state, H = Half-open-half-closed state and O = Open state

## IIIA. Rugged and smooth free energy surfaces of the LID and NMP domain motion

In **Figure 2b,** we show the free energy surfaces of the LID motion for intermediate NMP-CORE separation ($R^{CM}_{NMP-CORE} = 19$ Å). We find a stable minimum for the HOHC state of LID domain which has a $R^{CM}_{LID-CORE}$ around 26 Å. This again suggests the presence of an intermediate state of ADK. Fully open state is separated by a barrier of 20 kJ/mol and fully closed state is separated by a barrier of 25 kJ/mol from the HOHC state. Note that the free energy surface of LID domain motion is quite rugged. We will discuss this issue later in detail.

**Figure 2c** displays the free energy surfaces of the NMP domain motion for intermediate $R^{CM}_{LID-CORE}$ separations ($R^{CM}_{LID-CORE} = 25$ Å). We find that the minimum is located at $R^{CM}_{NMP-CORE} = 18.8$ Å. This is a stable intermediate state of ADK. Note that the free energy surface of NMP domain motion is relatively smooth as compared to the LID motion.

**Figure 2(b)**

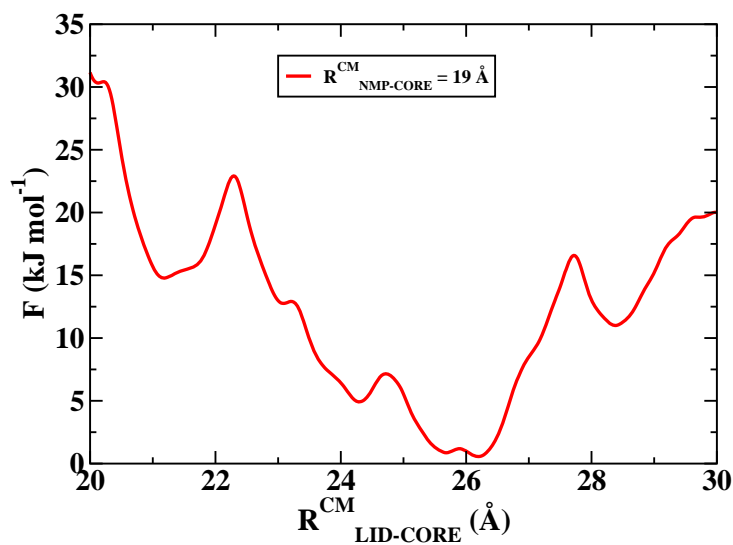

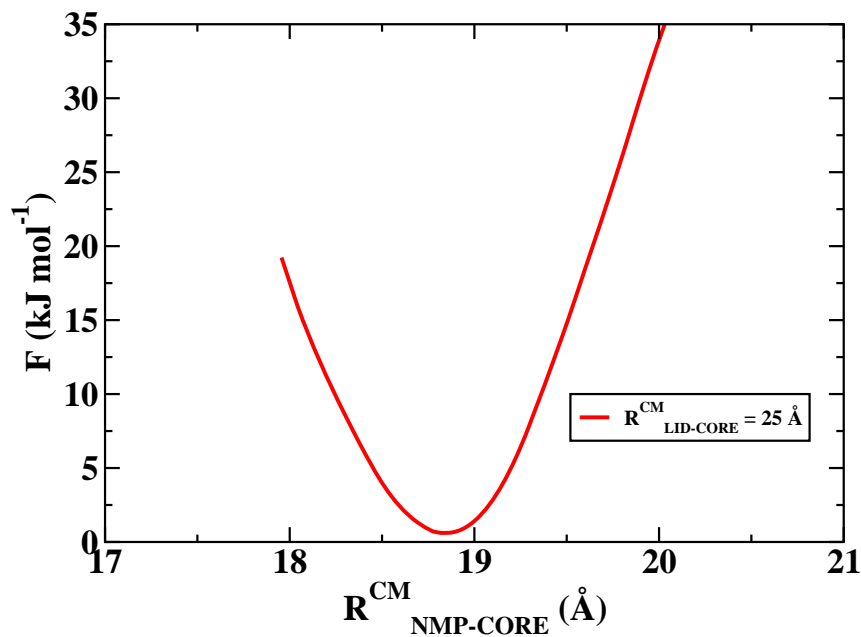

Figure 2(c)

## IV. Structural characterization of the HOHC state

In the intermediate HOHC state of the enzyme, the center of mass distance between LID and CORE domains lies between the open and closed states. We superimpose the structures of open (4AKE [27]), closed (1AKE [28]) and HOHC states of the enzyme to compare the domain positions as shown in **Figure 3**. The position of the LID domain in the HOHC state is clearly intermediate between the open and the closed states. The position of the NMP domain in the HOHC state is also found to be intermediate but closer to the open state structure.

A search for HOHC state in available crystal structures solved for ADK also reveals the presence of a naturally occurring stable intermediate structure (1AK2 [29]). *Superimposition of*

*the 1AK2 on closed and open forms of the enzyme indeed shows that the structure of 1AK2 corresponds to the structure of MD-proposed HOHC state in our study (see supplementary material S1).* In case of 1AK2, we also find that LID domain stays in an intermediate position, while NMP domain is in open state. The agreement is indeed startling!

**Figure 3**

**(a)**

**(b)**

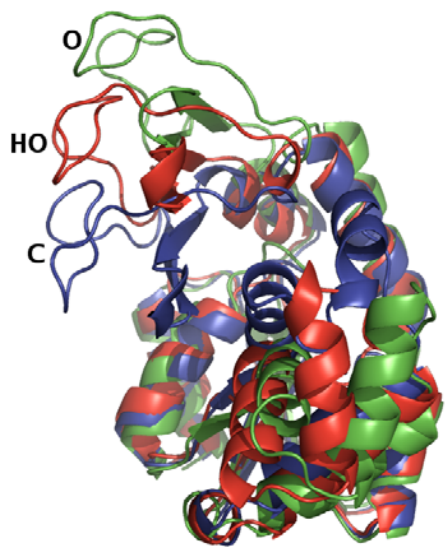
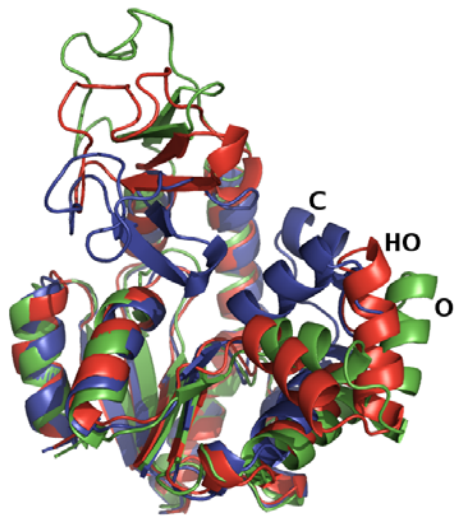

## V. Role of water in stabilizing the half-open-half-closed state

We have calculated the pair correlation function of the oxygen atoms of water with respect to the $C_\alpha$ atom of the polar histidine (His 134) residue, denoted by $g_{E,H2O}(r)$, and depicted in **Figure 4**. Note the appearance of the *new water layer* near the histidine residue in the HOHC state which *was not present in the closed state*. This result clearly indicates that upon partial opening of the ADK, the polar groups become solvated by the surrounding water and eventually stabilize the HOHC state of the enzyme.

**Figure 4**

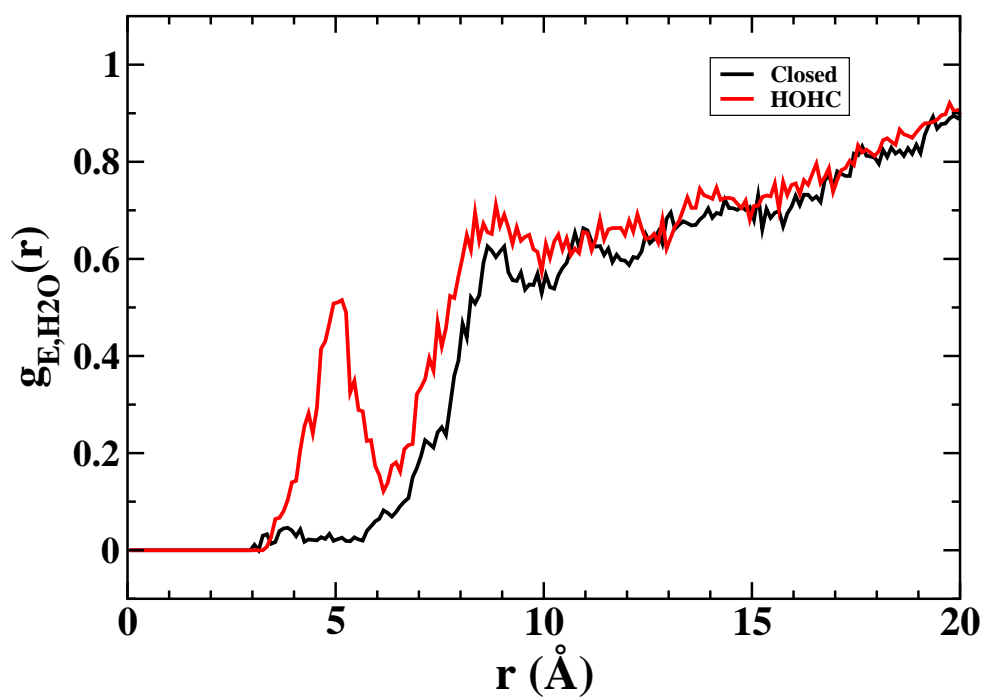

We have calculated the solvent accessible polar and non-polar surface area of the ADK in the closed and HOHC states. The surface representation of polar and non-polar residues of ADK and the calculated solvent accessible polar surface are in the two enzymatic states (see supplementary material **S2** for further formation) support the results obtained from $g_{E,H2O}(r)$. We also find a much larger increase in accessible polar surface area for the CORE-LID interface than

CORE-NMP interface upon opening which accounts for the more rugged free energy surface of the LID domain motion.

## VI. Role of the half-open-half-closed state in the catalytic cycle

A recent study on ADK has suggested that the ezymatic fluctuations are not important because the coupling between the enzyme fluctuation coordinate and the intrinsic reaction coordinate is rather weak [23, 30-32]. Recent simulation study by Brooks and cowerkers [22] suggests that the activation barrier from the open to the closed state is 52.3 KJ/mole (12.5 kcal/mol), leading to a timescale again in millisecond. Here we find quite a similar free energy barrier with the additional complexity due to the substantial ruggedness of the surface which will further slow down the open to closed state transition [33]. In this connection, there are two issues need to be considered. First, in a continuous catalytic cycle, the enzyme fluctuations might never operate from the fully open state [25-26]. *Thus, a continuous catalytic cycle might not experience the full free energy barrier for the transition from fully open state to the closed state* [25-26]. Therefore, in such a situation the ruggedness of the free energy surface of enzymatic fluctuation will enhance the rate of catalysis rather than retard it. Second is the involvement of the ATP and AMP in the transition state of the open to closed state transformation. For example, when LID is fully open, then the ATP binding to LID might actually be slower due to entropic reason, while ATP binding to the HOHC state might be faster.

Present study with its finding of the intermediate state suggests that the cycle would operate in its most efficient state if the cycle ranges from the closed to the HOHC state and back. This is illustrated in **Figure 5a** and the corresponding schematic contour diagram is presented in **Figure 5b**. If the enzyme captures ligand in HOHC state, its transition to closed state will be faster than that from completely open state [25-26] as discussed earlier in the discussion of the free two dimensional free energy surface section. This would reduce the time of the conformational cycling considerably. It is conceivable that both the conformational fluctuations and the phosphate transfer events have been tuned by the evolution to operate at certain rate. If the intervention of the HOHC state reduces the time required for the conformational cycling, then the intrinsic chemical reaction (phosphate transfer) step could be the rate determining step of the catalytic cycle. This would reconcile different theoretical observations. In figure 4b, we have presented schematic contour diagram of the different conformational states of ADK and the showed the important processes operating between them. Between HOHC and closed state the catalytic cycle operates and between HOHC and fully open state the experimentally observed millisecond conformational fluctuation occurs.

*A correlation observed between the opening of the LID and NMP domains suggests that the HOHC-LID state also facilitates the NMP to remain in the open form.* The free energy barrier for the subsequent transition to the closed state is further reduced by the induced fit mechanism (the interaction between the ligand and the enzyme) and the corresponding schematic free energy diagram [25-26] of the catalytic cycle is presented in supplementary material **S3**. The steady state rate (*v*) of catalysis can be decomposed into [26]

$$\frac{1}{v} = \frac{1}{k_{Chem}} + \frac{1}{k_{Conf}([S])} + \frac{1}{k_{S-bind}} \quad . \qquad \ldots\ldots\ldots\ldots (1)$$

Here $k_{Chem}$ is the intrinsic rate of the phosphate transfer step (ES → EP) (see **S3**), $k_{Conf}$ is the contribution due to enzyme fluctuations and $k_{S\text{-}bind}$ is the rate of substrate binding to the enzyme (can be given by Smoluchowski rate). It can be shown that under the non-equilibrium condition of a working enzyme, the barrier on the E....S surface (see Figure S3) can be reduced considerably at high substrate concentration and presence of stable intermediate on the E....S surface which can capture substrate [25-26]. Thus in such situation, $k_{E...S \to ES}$ (the step where enzyme fluctuation is important) can become large which in turn will make $k_{enz}$ large and essentially this term may dissappear from the above equation. Again, one can also argue that $k_{S\text{-bind}}$ can be large at the high substrate concentration limit, according Smoluchowski equation [25], and thus will make little contribution to the total rate. Thus in such a situation, the steady state will be determined only by the intrinsic rate of the reaction (phosphate stransfer step).

**Figure 5(a)**

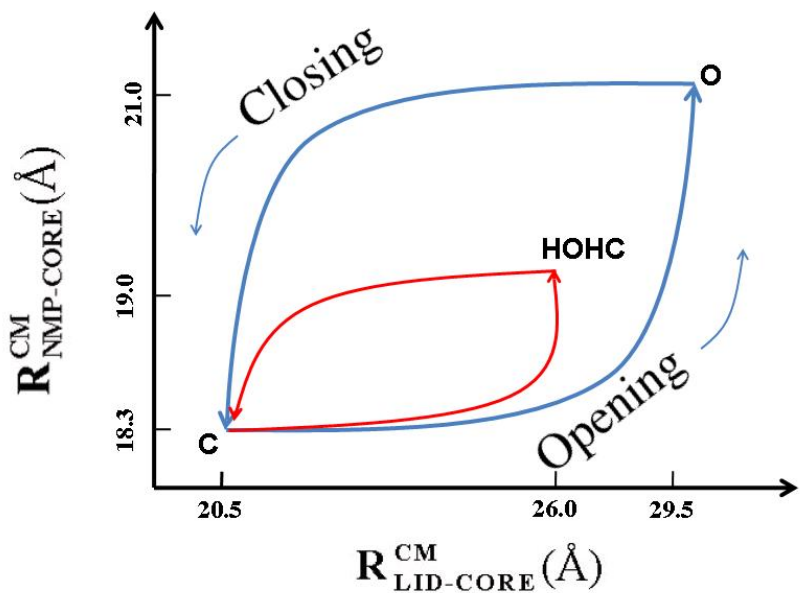

**Figure 5(b)**

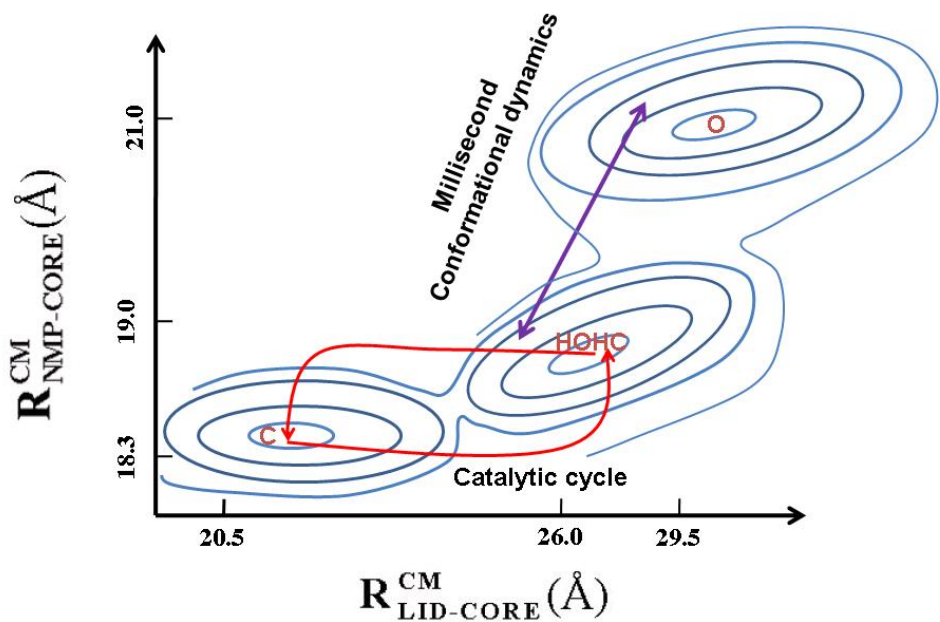

## VII. Conclusion

There are three new resuts in this paper. First, we observe and characterize, for the first time, the presence of a half-open-half-closed (HOHC) state during the sojourn of the closed form, after the product (ADP) release, towards the fully open state. The HOHC state has actually been implied in earlier NMR studies, but missed in earlier simulations that used only implicit solvent model. Our HOHC state is in excellent agreement with the naturally occuring stable intermediate IAK2. Our study reveals that the cycling of the enzyme through the HOHC state reduces the conformational free energy barrier by about 20 kJ/mol. Second, we find that the free energy surface of the closed form of the LID domain is quite rugged. This ruggedness is mainly due to the increased solvation by the surrounding water molecules upon partial opening. Finally, we propose a modfied catalytic cycle for continuous conversion (**Figure 5a**) and a corresponding free energy diagram of the catalytic cycle in the ezyme coordinate (supplementary material **S3**). A similar studies can also be carried out to understand the role of water in the emzymatic catalyic cycle of HIV-1 protease [15,24].

**System and Methods**

**Figure M1** shows the open and closed forms of the adenylate kinase superimposed on each other. The enzyme can be viewed as three domains: (a) LID domain: the domain which closes on ATP binding to the ADK, (b) NMP domain: the domain which closes on AMP binding to the ADK, and (c) CORE domain: the domain in which no significant conformational changes occurs upon ligand binding. The enzyme changes its conformation from open state to closed state during the ligand binding event. The LID and NMP domains open up to release the product and get ready to capture a new set of substrates for next round of catalysis. These two conformations can be characterized by the two distances [27,28]: (1) the distance between the centers of mass of the LID and the CORE domains, denoted by $R^{CM}_{LID-CORE}$ (2) the distance between the centers of mass of the NMP and the CORE domains, denoted by $R^{CM}_{NMP-CORE}$. The values of the $R^{CM}_{LID-CORE}$ for open and closed forms are 29.5 Å and 20.5 Å, respectively, and the same for the $R^{CM}_{NMP-CORE}$ for open and closed forms are 21.0 Å and 18.3 Å, respectively [27,28]. The LID domain is defined as residues 118-160, the NMP domain as residues 30-67, and the CORE domain as residues 1-29, 68-117, and 161-214.

As mentioned earlier, several studies have been carried out to understand the role of conformational fluctuations on catalysis both experimentally and computationally. Here, we report for the first time a very long time scale (individual runs of maximum 500 ns and total run time 2 µs) fully atomistic simulation of ADK in explicit solvent environment. The long time scale provides us a unique opportunity to comment confidently on conformational fluctuations and probable binding events that might be happening in ADK.

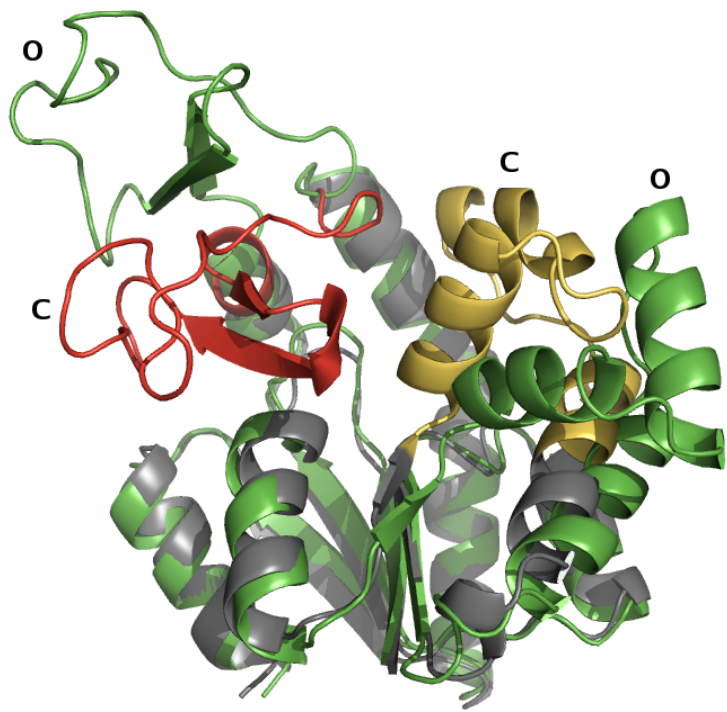

**Figure M1:** **Ribbon diagram open and closed forms of the adenylate kinase**. Superposition of closed (1AKE [28]) and open (4AKE [27]) forms of adenylate kinase. The open form is colored green. The LID domain in closed form is colored in red, the NMP domain is in yellow, and the CORE domain is in gray.

The x-ray structure of the open form (4AKE [27]) was used as the starting structure for open state simulations. To generate the starting structure for the closed state simulation, ligand (inhibitor AP5) coordinates were removed from the x-ray structure of the closed form (1AKE [28]). All crystallographic water molecules were stripped off from the pdb files. GROMACS v3.3.1 suite of programs was used for molecular dynamics and other structural analysis [34]. Proteins were centered in a cubic box of 77.3 A. The box size was so chosen that no atoms in the protein, either in open or in closed conformation, will be less than 10 Å from any of the box-boundaries. All atom topologies for proteins were generated with the help of pdb2gmx and AMBER94 set of parameters (available through AMBER port for GROMACS).

The proteins were solvated with pre-equilibrated SPC/E water model [35] using genbox. Total of 14511 and 14496 water molecules were added to closed and open state boxes, respectively. Four sodium atoms were added to achieve electrically neutrality. The systems were put through following cassette of energy minimization and equilibration steps: 1) Steepest descent energy minimization, 2) A 100 ps of position-restrained NPT simulation with restraining force constant of 1000 kJ.mol$^{-1}$nm$^{-2}$, 3) A 100 ps of NVT equilibration without restraints, 4) A 100 ps of NPT equilibration, 5) The final NPT simulation of 350 ns. During simulations, temperature was maintained at 300 K using Nose-Hoover thermostat with $\tau_T = 0.1$ ps and pressure was maintained isotropically at 1 bar using Berendsen barostat with $\tau_P = 0.5$ ps. Isothermal compressibility of water was set to $4.5 \times 10^{-5}$ bar$^{-1}$. The trajectories are built with structures at 2 ps interval.

**RMSD:** Root mean square deviation for backbone was calculated with the help of g_rms program when backbone of CORE domain was used for superimposition. The structures from both the trajectories were superimposed onto the closed conformation of ADK. A common reference for both the trajectories was chosen so that the distribution of RMSD can be easily studied. LID domain is defined as residues 118-160, NMP as 30-67, and CORE domain is defined as residues 1-29, 68-117, and 161-214.

**Angle and distance:** The Cα atoms of following residues are used to calculate distance between LID and CORE domains with g_dist: Val148 (from LID) and Glu22 (from CORE). The angle calculation is done by considering Cα atom of Pro9 (from hinge) in addition to Cα from Val148 and Glu22 with g_angle.

**Umbrella sampling:** While $R^{CM}_{LID-CORE}$ varies from 20.5 Å to 29.5 Å while going from closed conformation to open conformation, $R^{CM}_{NMP-CORE}$ varies from 18.3 Å to 21.5 Å. A two-dimensional free-energy surface was constructed by taking $R^{CM}$ as reaction coordinates. $R^{CM}_{LID-CORE}$ varied from 19.0 Å to 32 Å, and $R^{CM}_{NMP-CORE}$ from 17 Å to 23 Å, both at an interval of 0.5 Å. Force constant of 3000 kJ.mol$^{-1}$.nm$^{-2}$ was used to restrain the domains at respective $R^{CM}$. Umbrella sampling simulations were performed using pull code from GROMACS v4.0.5. Free energy surfaces are extracted using wham program. The simulations involved more than 80,000 water molecules for the free energy calculations.

ACKNOWLEDGEMENT

We thank Dr. H. Balaram, Dr. P. Balaram, Dr. J. Klinman, Dr. A. Warshel and Dr. P. Wolynes for helpful discussions. This work was supported in parts by grants from DST (India) and CSIR (India). BB was supported by a J.C. Bose Fellowship. BJ thanks CSIR for a Fellowship.

**Supporting Information Available**

(1) Experimental structure of 1AK2 obtained from ref. 29 is compared with the HOHC state obtained from simulation. (2) Surface representation of the HOHC state of ADK to show the exposed polar and non-polar surfaces. (3) Modified free energy surface of the chmical reaction by ADK in the presence of HOHC state. This material is available free of charge via the Internet at http://pubs.acs.org.

# Figure Captions

**Figure 1: Comparison of fluctuations to confirm the inter-domain nature of the movements**. The RMSD of LID domain with respect to the native structure of the closed state, LID-hinge-CORE angle, and LID-CORE distance distribution as a function of simulation time for closed state simulation. The Cαs of following residues are used to calculate distance between LID and CORE domains: Val148 (from LID) and Glu22 (from CORE). The angle calculation is done by considering Cα atom of Pro9 (from hinge) in addition to Cα from Val148 and Glu22. Note the similarity is the fluctuations of the different quantities as highlighted by the selected region of ellipses. The region highlighted by the red ellipse shows the fluctuation in the HOHC state.

**Figure 2: (a) Two-dimensional free energy surface of the free ADK enzyme.** Two dimensions are the center of mass distances of the LID and NMP domains from the CORE. The color code has been so chosen that the closely spaced regions can be distinguished clearly. Note the stable region around HOHC state (blue region). **(b) Free energy landscape of the LID domain motion.** Free energy surface calculated from umbrella sampling for the LID domain motion at $R^{CM}_{NMP-CORE} = 19$ Å. Note the minima at HOHC state. Note also the rugged landscape of the free energy surface. **(c) Free energy

**landscape of the NMP domain motion.** Free energy surface calculated from umbrella sampling for the NMP domain motion at $R^{CM}_{LID\text{-}CORE} = 25$ Å. Note the minimum about 19 Å indicating the stability of HOHC state and also the relatively smooth landscape.

**Figure 3: Structure of the HOHC state observed in simulations.** Superposition of closed (1AKE [28]) and open (4AKE [27]) states of ADK along with HOHC state. LID and NMP domains are shown as cartoon. Closed state is colored in green, open state in blue, and HOHC state is colored in orange color. **a)** The view shows the intermediate position of the LID domain in HOHC state. **b)** 90˚ rotation about y-axis to show alignment of the NMP domain. Here HO indicates the HOHC state.

**Figure 4: Stabilization of the HOHC by water.** Pair correlation functions of water molecules with respect to the $C_\alpha$ atom of the polar histidine residue in the HOHC and the closed state of the enzyme. Note the appearance of a new layer of water (~ 5 Å) in the HOHC state of the enzyme.

**Figure 5: Modified catalytic cyale of the chmical reaction by ADK and schematic contour digram.** (a)Full cycle shows the catalytic process involving closed to fully open state of the ADK. The cycle involving red line shows the modified catalytic cycle involving HOHC state. (b) Schematic contour digram has been presented for the different conformational states ADK. The catalytic cycle (red lines) involves closed and HOHC states. The millisecond conformational fluctuation observed in experiment involves fluctuation between HOHC and fully open state.

# Table of Content Graphic

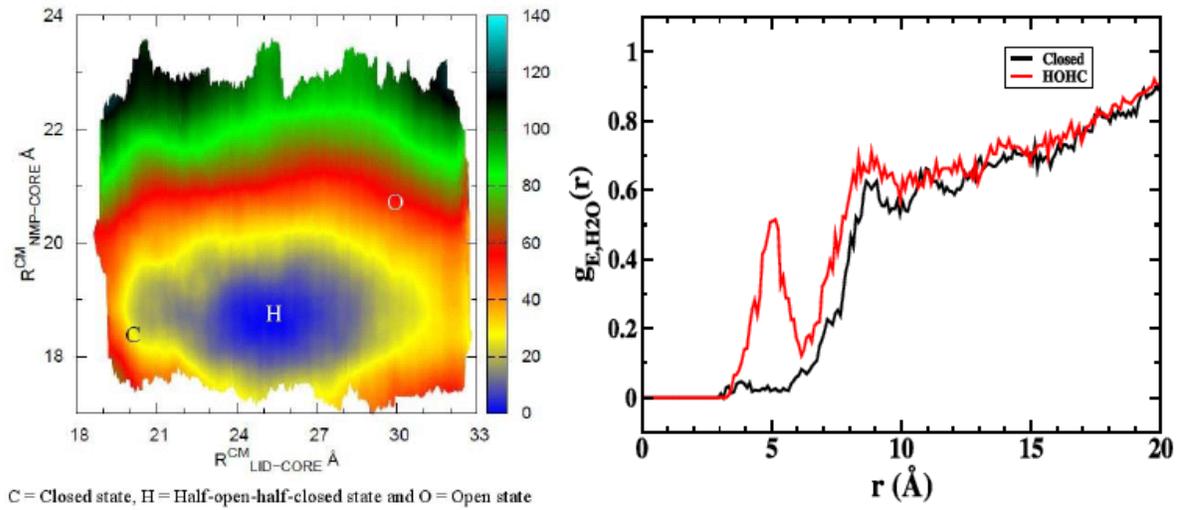

C = Closed state, H = Half-open-half-closed state and O = Open state

# Supplementary Information

**S1:**

**(a)**                                    **(b)**

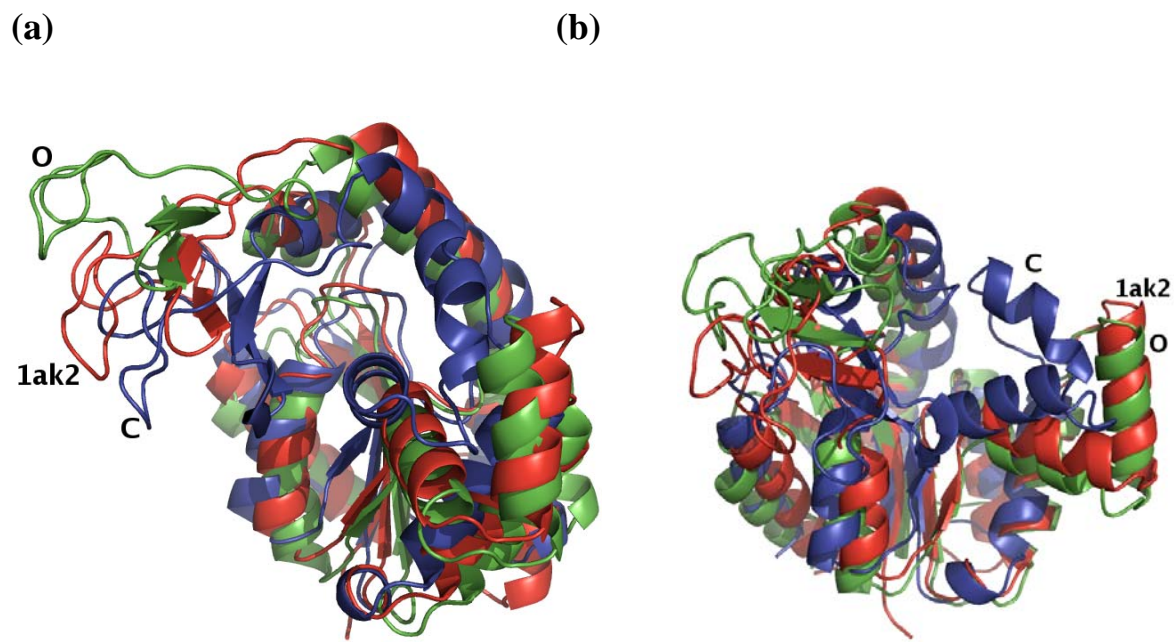

**Figure S1: Experimental structures of 1AK2 obtained from ref. 29 to be compared with the HOHC state obtained from simulation.** Superposition of 1AK2 [29] structure (red) with closed (1AKE [28]) (blue) and open (4AKE [27]) (green) state structures is at an angle where **a)** intermediate conformation of LID domain can be viewed properly and where **b)** intermediate conformation of NMP domain can be viewed properly.

## S2:

Surface representation of the HOHC state of ADK is presented here (**Figure S2)**. We find that the polar amino residues (represented by light colors of the corresponding domain color) are becoming exposed in the inter-domain region. These polar amino acids interact with water molecules and eventually water mediated interactions increase. This increase in inter-domain water mediated interactions upon partial opening of the domains is reflected in the pair correlation function $g_{E,H2O}(r)$ as shown in **Figure 4.**

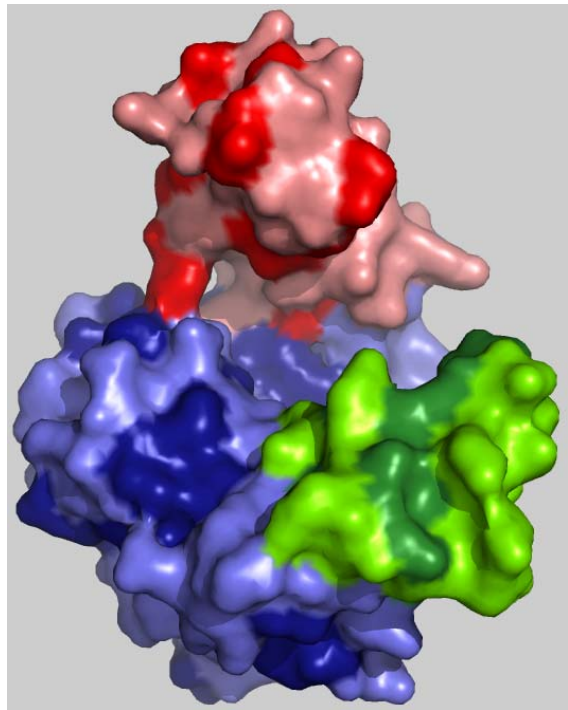

**Figure S2: Surface representation of the HOHC state of ADK.** The domains are in different colors – LID in red, NMP in green and CORE in blue. The polar residues in each domain are colored in respective light color. Note the increase in the accessible polar surface area in the inter domain region, in particular for CORE-LID region, in the HOHC state.

Estimation of polar and non-polar surface area in the closed and HOHC states inferred that while the non-polar surface area remains almost similar in both the structures, polar surface area increases significantly in the HOHC state compared to the closed state. This indicates the increase of the water mediated interaction in the interface region upon domain opening of the ADK.

**S3:**

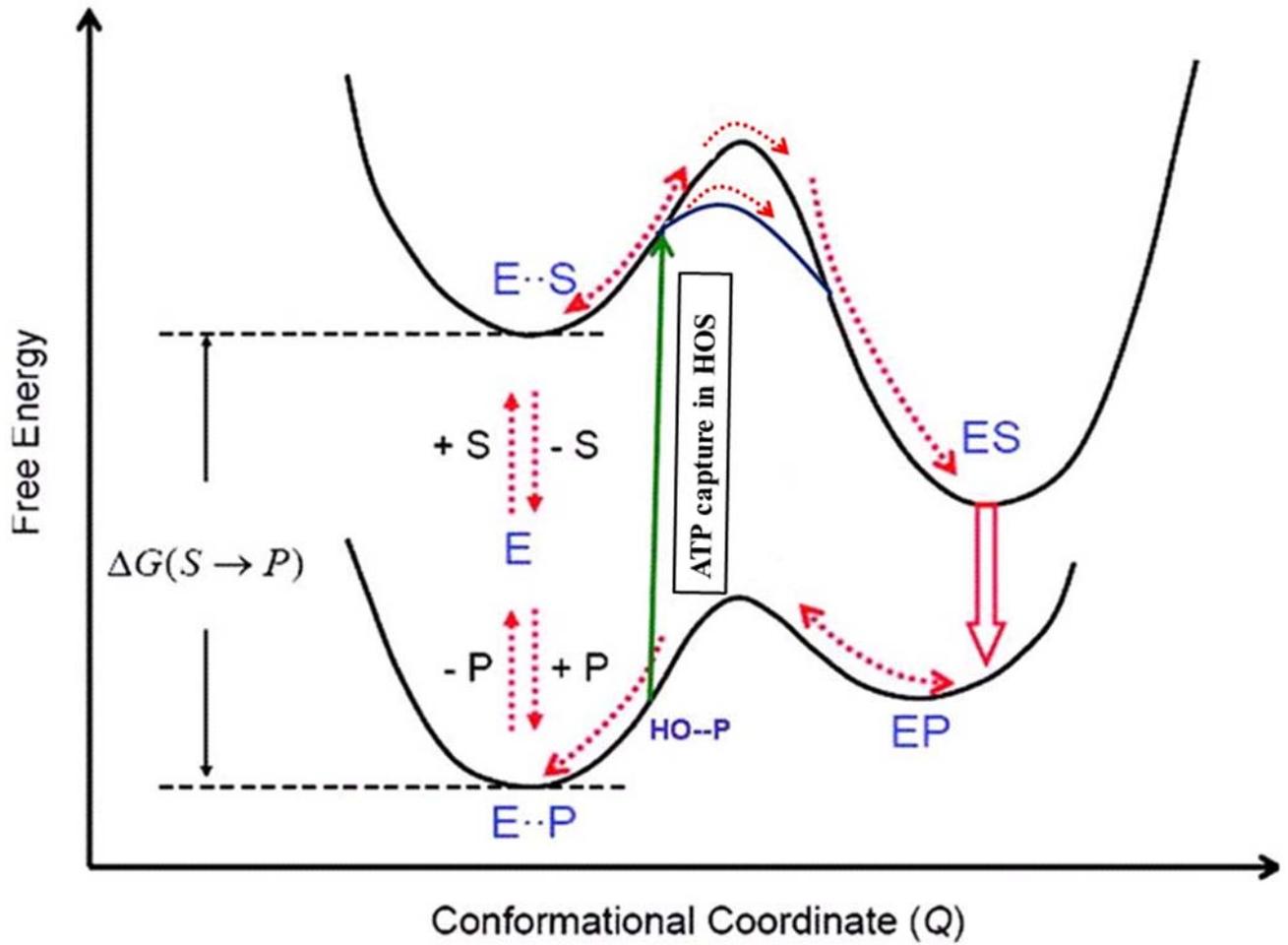

**Figure S3: Free energy surface of the chemical reaction by ADK.** Corresponding free energy surface of the modified cycle. Green line shows the conversion of E…..S complex from HO---P state (here HO indicates the HOHC state). Deep blue line show the corresponding facilitate induced fit of the E….S complex (from HO----P complex) with low free energy barrier.